# A $3\omega$ method for specific heat and thermal conductivity measurements

L. Lu[*], W. Yi and D. L. Zhang

*Laboratory of Extreme Conditions Physics, Institute of Physics & Center for Condensed Matter Physics, Chinese Academy of Sciences, Beijing 100080, P. R. China*

We present a $3\omega$ method for simultaneously measuring the specific heat and thermal conductivity of a rod- or filament-like specimen using a way similar to a four-probe resistance measurement. The specimen in this method needs to be electrically conductive and with a temperature-dependent resistance, for acting both as a heater to create a temperature fluctuation and as a sensor to measure its thermal response. With this method we have successfully measured the specific heat and thermal conductivity of platinum wire specimens at cryogenic temperatures, and measured those thermal quantities of tiny carbon nanotube bundles some of which are only $\sim 10^{-9}$ g in mass.

## I. INTRODUCTION

Many experimental methods have been developed over the past centuries to measure the fundamental thermal properties of materials. One important class among them, the so called $3\omega$ method, uses a narrow-band detection technique and therefore gives a relatively better signal-to-noise ratio. In this method, either the specimen itself serves as a heater and at the same time a temperature sensor, if it is electrically conductive and with a temperature-dependent electric resistance, or for electrically non-conductive specimen, a metal strip is artificially deposited on its surface to serve both as the heater and the sensor. Feeding an *ac* electric current of the form $I_0\sin\omega t$ into the specimen or the metal strip creates a temperature fluctuation on it at the frequency $2\omega$, and accordingly a resistance fluctuation at $2\omega$. This further leads to a voltage fluctuation at $3\omega$ across the specimen. Corbino[1] is probably the first to notice that the temperature fluctuation of an *ac* heated wire gives useful information about the thermal properties of the constituent material. Systematic investigations of the $3\omega$ method were carried out mainly during the 1960's[2–4] and in the last ten years[5–10], which made the method practicable. However, in the previous studies the heat-conduction equation was solved under the approximations either only for the high frequency limit[2,3,10], or only for the low frequency limit[5,7,8]. With those approximations one lost either the information on the thermal conductivity or the information on the specific heat of the specimen.

In this paper we present an explicit solution for the 1D heat-conduction equation. With this solution and by using a modern digital lock-in amplifier, we are able to obtain both the specific heat and the thermal conductivity of a rod- or filament-like specimen simultaneously. We have tested this method on platinum wire specimens. Correct values of specific heat, thermal conductivity, and Wiedemenn-Franz ratio were obtained. With this method we have also obtained the thermal properties of carbon nanotube bundles some of which are only $10^{-9}$ g in mass.

In section II we will present an explicit solution for the 1D heat-conduction equation. In section III we will discuss the high and low frequency limits of the solution, then comparing them with the ones previously obtained by others at these limits. An error analysis will be given in section IV, for the case of just keeping the first term of the solution. In section V we will discuss the effects of radial heat loss. And in section VI we will show our experimental test of the method on platinum and carbon nanotube materials. We will also share with the readers the tips of using this $3\omega$ method.

## II. THE 1D HEAT CONDUCTION EQUATION AND ITS SOLUTION

We consider a uniform rod- or filament-like specimen in a four-probe configuration as for electrical resistance measurement (Fig. 1). The two outside probes are used for feeding an electric current, and the two inside ones for measuring the voltage across the specimen. Differing from being a pure electrical resistance measurement, however, here it requires that (i) the specimen in-between the two voltage probes be suspended, to allow the temperature fluctuation. (ii) All the probes have to be highly thermal-conductive, to heat-sink the specimen at these points to the sapphire substrate. (iii) The specimen has to be maintained in a high vacuum and the whole setup be heat-shielded to the substrate temperature, to minimize the radial heat loss through gas convection and radiation. In such a configuration and with an *ac* electrical current of the form $I_0\sin\omega t$ passing through the specimen, the heat generation and diffusion along the specimen can be described by the following partial differential equation and the initial and boundary conditions:

$$\rho C_p \frac{\partial}{\partial t} T(x,t) - \kappa \frac{\partial^2}{\partial x^2} T(x,t) = \frac{I_0^2 \sin^2\omega t}{LS}[R + R'\left(T(x,t) - T_0\right)] \quad (1)$$

$$\begin{cases} T(0,t) = T_0 \\ T(L,t) = T_0 \\ T(x,-\infty) = T_0 \end{cases} \quad (2)$$

where $C_p$, $\kappa$, $R$ and $\rho$ are the specific heat, thermal conductivity, electric resistance and mass density of the specimen at the substrate temperature $T_0$, respectively. $R' = \left(\frac{dR}{dT}\right)_{T_0}$. $L$ is the length of the specimen between voltage contacts, and $S$ the cross section of the specimen. We have assumed that the electric current was turned on at $t = -\infty$.

Let $\Delta(x,t)$ denote the temperature variation from $T_0$, i.e., $\Delta(x,t) = T(x,t) - T_0$, equations (1) and (2) then become:

$$\frac{\partial}{\partial t}\Delta(x,t) - \alpha\frac{\partial^2}{\partial x^2}\Delta(x,t) - c\sin^2\omega t\ \Delta(x,t) = b\sin^2\omega t \tag{3}$$

$$\begin{cases} \Delta(0,t) = 0 \\ \Delta(L,t) = 0 \\ \Delta(x,-\infty) = 0 \end{cases} \tag{4}$$

where $\alpha = \frac{\kappa}{\rho C_p}$ is the thermal diffusivity, $b = \frac{I_0^2 R}{\rho C_p LS}$, $c = \frac{I_0^2 R'}{\rho C_p LS}$.

Using the impulse theorem, $\Delta(x,t)$ can be represented as the integral of specimen's responses to the instant "force" $b\sin^2\omega t$ at each time interval:

$$\Delta(x,t) = \int_{-\infty}^{t} z(x,t;\tau)d\tau \tag{5}$$

where $z(x,t;\tau)$ satisfies:

$$\frac{\partial z}{\partial t} - \alpha\frac{\partial^2 z}{\partial x^2} - c\sin^2\omega t\ z = 0 \tag{6}$$

$$\begin{cases} z(0,t) = 0 \\ z(L,t) = 0 \\ z(x,\tau+0) = b\sin^2\omega\tau \end{cases} \tag{7}$$

$z(x,t;\tau)$ can be expanded in the Fourier series:

$$z(x,t;\tau) = \sum_{n=1}^{\infty} U_n(t;\tau)\sin\frac{n\pi x}{L} \tag{8}$$

Substituting (8) into (6), we have

$$\sum_{n=1}^{\infty}\left[\frac{dU_n}{dt} + \left(\frac{n^2}{\gamma} - c\sin^2\omega t\right)U_n\right]\sin\frac{n\pi x}{L} = 0 \tag{9}$$

where $\gamma \equiv L^2/\pi^2\alpha$ is the specimen's characteristic thermal time constant for axial thermal process.

The term $c\sin^2\omega t$ can be neglected if $n^2/\gamma \gg c$, or equivalently

$$\frac{I_0^2 R' L}{n^2\pi^2\kappa S} \ll 1 \tag{10}$$

Condition (10) means that the heating power inhomogeneity caused by resistance fluctuation along the specimen should be much less than the total heat power. This condition is usually held. For example, in a typical measurement one could have $I_0$=10 mA, $R'$=0.1 $\Omega$/K, $L$=1 mm, $S$=$10^{-2}$ mm$^2$, $\kappa$=100 W/m K, the left side of (10) is then about $10^{-3}$ even for the $n=1$ case.

After dropped off the $c\sin^2\omega t$ term, the solution of the ordinary differential equation (9) is:

$$U_n(t;\tau) = C_n(\tau)e^{-\frac{n^2}{\gamma}(t-\tau)} \tag{11}$$

where $C_n(\tau)$ can be determined using the initial condition in (7), together with the relation $\sum_{n=1}^{\infty}\frac{2[1-(-1)^n]}{n\pi}\sin\frac{n\pi x}{L} = 1$ for $0 < x < L$:

$$C_n(\tau) = \frac{2b[1-(-1)^n]}{n\pi}\sin^2\omega\tau \tag{12}$$

Using (11) and (12), (8) becomes:

$$z(x,t;\tau) = \sum_{n=1}^{\infty}\sin\frac{n\pi x}{L}\ \frac{2b[1-(-1)^n]}{n\pi}\ \sin^2\omega\tau\ e^{-\frac{n^2}{\gamma}(t-\tau)} \tag{13}$$

Substituting (13) into (5) and remembering that $\Delta(x,t) = T(x,t) - T_0$, we obtain the temperature distribution along the specimen:

$$T(x,t) - T_0 = \Delta_0\sum_{n=1}^{\infty}\frac{[1-(-1)^n]}{2n^3}\sin\frac{n\pi x}{L}\left[1 - \frac{\sin(2\omega t+\phi_n)}{\sqrt{1+\cot^2\phi_n}}\right] \tag{14}$$

where $\cot\phi_n = \frac{2\omega\gamma}{n^2}$, and $\Delta_0 = \frac{2\gamma b}{\pi} = \frac{2I_0^2 R}{\pi\kappa S/L}$ is the maximum $dc$ temperature accumulation at the center of the specimen. $\Delta_0$ is only $\kappa$-dependent. The information of $C_p$ is included in the fluctuation amplitude of the temperature around the $dc$ accumulation.

Figure 2 illustrates how the amplitude of such temperature fluctuation depends on the frequency of the electric current. The amplitude reaches the maximum as $\omega\gamma \to 0$, i.e., when the thermal wavelength $\lambda \gg L$ (where $\lambda$ is defined as $\lambda = \sqrt{\frac{\alpha}{2\omega}}$). But it shrinks to zero along the line of the averaged temperature accumulation when $\omega\gamma \gg 1$ ($\lambda \ll L$).

The temperature fluctuation will result in a resistance fluctuation, which can be calculated as:

$$\delta R = \frac{R'}{L}\int_0^L [T(x,t) - T_0]dx \tag{15}$$

Using (14) and the relation $\int_0^L \sin\frac{n\pi x}{L}dx = [1-(-1)^n]\frac{L}{n\pi}$, the resistance fluctuation can be expressed as:

$$\delta R = R'\Delta_0\sum_{n=1}^{\infty}\frac{[1-(-1)^n]^2}{2\pi n^4}\left[1 - \frac{\sin(2\omega t+\phi_n)}{\sqrt{1+\cot^2\phi_n}}\right] \tag{16}$$

As a product of the total resistance $R+\delta R$ and the current $I_0\sin\omega t$, the voltage across the specimen contains a $3\omega$ component $V_{3\omega}(t)$. Obviously, the $n = 2$ term in $V_{3\omega}(t)$ automatically vanishes. If only taking the $n = 1$ term, which introduces a relative error of the order $\sim 3^{-4}$ at low frequencies, we have:

$$V_{3\omega}(t) \approx -\frac{2I_0^3 L R R'}{\pi^4 \kappa S \sqrt{1+(2\omega\gamma)^2}} \sin(3\omega t - \phi) \quad (17)$$

where we have re-defined the phase constant $\phi = \frac{\pi}{2} - \phi_1$ so that:

$$\tan\phi \approx 2\omega\gamma \quad (18)$$

If using the rms values of voltage and current as what lock-in amplifier gives, equation (17) becomes (hereafter we always let $V_{3\omega}$ denotes the rms value of $V_{3\omega}(t)$, and $I$ denotes the rms value of $I_0 \sin\omega t$ ):

$$V_{3\omega} \approx \frac{4 I^3 L R R'}{\pi^4 \kappa S \sqrt{1+(2\omega\gamma)^2}} \quad (19)$$

By fitting the experimental data to this formula we can get the thermal conductivity $\kappa$ and thermal time constant $\gamma$ of the specimen. The specific heat can then be calculated as:

$$C_p = \pi^2 \gamma \kappa / \rho L^2 \quad (20)$$

The following alternative form makes it more clear how the $3\omega$ voltage depends on the dimensions of the specimen:

$$V_{3\omega} \approx \frac{4 I^3 \rho_e \rho'_e}{\pi^4 \kappa \sqrt{1+(2\omega\gamma)^2}} \left(\frac{L}{S}\right)^3 \quad (21)$$

where $\rho_e$ is the electrical resistivity of the specimen, $\rho'_e \equiv (d\rho_e/dT)$.

## III. THE HIGH AND LOW FREQUENCY LIMITS

Sometime the measurement has to be performed at the low frequency limit $\omega\gamma \to 0$ ($\lambda \gg L$), *e.g.*, when the specimen is extremely thin and long. In this case $V_{3\omega}$ is nearly frequency-independent. To an accuracy of roughly $3^{-4}$, it takes the form:

$$V_{3\omega} \approx \frac{4 I^3 R R' L}{\pi^4 \kappa S} = \frac{1}{\pi^3} I R' \Delta_0 \quad (\omega\gamma \to 0) \quad (22)$$

If the measurement is performed at the low frequency limit, one can only get the thermal conductivity of the specimen, but loses the information on specific heat, as in Cahill's treatment for a two-dimensional heat diffusion problem[8].

At the high frequency limit $\omega\gamma \to \infty$ ($\lambda \ll L$), on the other hand, equations (17) to (21) become quite inaccurate due to truncating the $n > 1$ terms in (16). In this limit, all the $\phi_n$ approach to zero, and the amplitude of the summation over the time-dependent terms in (16) eventually becomes $\sum_{n=1,\text{odd}}^{\infty} n^{-2} = \pi^2/8$. Therefore, $V_{3\omega}$ should be:

$$V_{3\omega} = \frac{I^3 R R'}{4\omega\rho C_p L S} \quad (\omega\gamma \to \infty) \quad (23)$$

which is exactly the same as Holland's result[2]. Simply truncating the $n > 1$ terms at the $\omega\gamma \to \infty$ limit will result in a coefficient of $2/\pi^2$, instead of $1/4$, for $V_{3\omega}$ in (23).

At the high frequency limit, one can only get the specific heat of the specimen, but loses the information on its thermal conductivity.

## IV. ERROR ANALYSIS

The error of $V_{3\omega}$ caused by truncating the $n > 1$ terms in (16) is illustrated in Fig. 3. Curve A is the normalized fluctuation amplitude of (16) containing terms up to $n = 9$, taken from a numerically generated time sequence. It almost represents the exact solution. Curve B is the fluctuation amplitude of the first term alone. It appears that the difference between A and B (shown as curve A-B in Fig. 3) is nearly a constant in the frequency range $0 < 2\omega\gamma < 10$. It approaches to $\sum_{n=3,\text{odd}}^{\infty} n^{-4} \approx 0.014$ as $\omega \to 0$. However, because $V_{3\omega}$ decreases with frequency, the relative error of $V_{3\omega}$ increases with $\omega$ (illustrated as curve (A-B)/A in Fig. 3).

The relative error of $\tan\phi$ in (18) should also increase with frequency. Indeed, the experimental data of $\tan\phi$ do curve away from linearity at high frequencies. By fitting the data to (18), the high-frequency inaccurate side of (18) provides more weight on the slope, so that one will get a noticeably smaller $\gamma$ than the true value.

The case of using (19) is fortunately just the opposite. The amplitude of $V_{3\omega}$ is relatively large at the low frequency side where (19) is very accurate. If we fit curve A to (19) in the frequency range $0 < 2\omega\gamma < 4$, the obtained $\kappa$ is only 3.5% higher, and $\gamma$ 2% lower than the true values. $C_p$ is then only 1.4% higher than the true value.

Because the error in (19) is nearly frequency independent at low frequencies (curve A-B in Fig. 3), it can be further and easily reduced by shifting the fitting curve upwards by a small amount, *i.e.*, fitting the data to the following form[11]:

$$V_{3\omega} \approx \frac{4 I^3 L R R'}{1.01 \pi^4 \kappa S} \left[\frac{1}{\sqrt{1+(2\omega\gamma)^2}} + 0.01\right] \quad (2\omega\gamma \le 4) \quad (24)$$

Fitting curve A to (24) in the frequency range $0 < 2\omega\gamma < 4$ yields $\kappa$, $\gamma$ and $C_p$ that are all within 0.1% of their true values. In this case the error introduced by truncating the $n < 1$ terms becomes negligibly small comparing with the errors of other sources, such as from the size estimation.

If one truncates the $n > 1$ terms in (14) to calculate the temperature fluctuation, the error will be more significant than truncating the $n > 1$ terms in (16). This is because the summation converges as $n^{-3}$ in (14), not as $n^{-4}$ in (16).

## V. RADIAL HEAT LOSS

In the above we have neglected the radial heat loss through radiation. Such heat loss per unit length from a cylindrical rod of diameter $D$ to the environment of temperature $T_0$ can be expressed as:

$$W(x,t) = \pi\epsilon\sigma D\left[T^4(x,t) - T_0^4\right] \approx 4\pi\epsilon\sigma D T_0^3 \Delta(x,t) \tag{25}$$

where $\sigma = 5.67\times 10^{-8}$ W/m$^2$K$^4$ is the Stefan-Boltzmann constant, and $\epsilon$ is the emissivity.

Considering such heat loss, (3) and (4) can be rewritten as:

$$\frac{\partial}{\partial t}\Delta(x,t) - \alpha\frac{\partial^2}{\partial x^2}\Delta(x,t) + \left(g - c\sin^2\omega t\right)\Delta(x,t) = b\sin^2\omega t \tag{26}$$

$$\begin{cases} \Delta(0,t) = 0 \\ \Delta(L,t) = 0 \\ \Delta(x,-\infty) = 0 \end{cases} \tag{27}$$

where $g = \frac{16\epsilon\sigma T_0^3}{\rho C_p D}$. Equation (9) then becomes:

$$\sum_{n=1}^{\infty}\left[\frac{dU_n}{dt} + \left(\frac{n^2}{\gamma} + g - c\sin^2\omega t\right)U_n\right]\sin\frac{n\pi x}{L} = 0 \tag{28}$$

Now if we truncate the $n > 1$ terms again and replace the factor $\frac{1}{\gamma} + g$ with $\frac{1}{\gamma_{ap}}$, equation (28) will take the similar form as (9). The final approximation solution is therefore:

$$V_{3\omega} \approx \frac{4I^3 LRR'}{\pi^4 S\kappa_{ap}\sqrt{1+(2\omega\gamma_{ap})^2}} \tag{29}$$

$$\tan\phi \approx 2\omega\gamma_{ap} \tag{30}$$

where $\kappa_{ap} = (1+g\gamma)\kappa$ is the apparent thermal conductivity, and $\gamma_{ap} = \gamma/(1+g\gamma)$ is the apparent thermal time constant of the specimen. The apparent $dc$ temperature accumulation is $\Delta_0^{ap} = \Delta_0/(1+g\gamma)$ at the center of the specimen.

Obviously, radiation heat loss can be neglected if

$$g\gamma \ll 1 \tag{31}$$

For cylindrical rod, condition (31) becomes $\frac{16\epsilon\sigma T_0^3 L^2}{\pi^2\kappa D} \ll 1$, which means that the radiation power inhomogeneity caused by the temperature fluctuation along the specimen should be much less than the axial heat current or the total heating power.

Condition (31) is usually held for measurements performed below room temperature. For example, if one has a specimen of the size $L$=1 mm, $D$=10$^{-2}$ mm, and assuming $\kappa$=100 W/m K, $T_0$=300 K, the product $g\gamma$ is only around $2.5\times 10^{-3}$ even if using the emissivity of a black body.

However, for specimens of significantly longer or thinner, or if the measurement is performed at significantly higher temperatures, condition (31) will be violated. In these cases the apparent thermal conductivity is larger than the actual value by an amount due to the radial heat loss, for cylindrical rod which is:

$$\kappa_{ap} = \kappa(1+g\gamma) = \kappa + \frac{16\epsilon\sigma T_0^3 L^2}{\pi^2 D} \tag{32}$$

If one knows the emissivity, then both $\kappa$ and $C_p$ of the specimen can be calculated. Otherwise if the emissivity is unknown, one will lose the information of $\kappa$. Nevertheless, one can still get $C_p$ of the specimen. The reason is, by substituting $\kappa_{ap}$ and $\gamma_{ap}$ into (20) as if there is no radial heat loss, the $(1+g\gamma)$ factors in $\kappa_{ap}$ and in $\gamma_{ap}$ just cancels out, which yields the correct value of $C_p$:

$$C_p = \pi^2\gamma_{ap}\kappa_{ap}/\rho L^2 \equiv \pi^2\gamma\kappa/\rho L^2 \tag{33}$$

Although the above analysis is made for cylindrical rod, the conclusions are also revelatory for specimens of other shapes. One can easily deduce the factor $g\gamma$ for particular specimens if needed.

Another kind of radial heat loss, the heat loss through gas convection, also introduces a linear-term correction to the heat conduction equation. The final solution is therefore the same as (29) and (30) except that now $g = \frac{4\eta}{\rho C_p D}$ for cylindrical specimen of diameter $D$ (where $\eta$ is the surface thermal conductivity). Similar to the case of radiation heat loss, one need to know $\eta$ before being able to calculate $\kappa$. But one can still obtain $C_p$ of the specimen through (33) without knowing $\eta$. This has been proven to be true experimentally, even when the heat loss through gas convection is much larger than the axial thermal current (the experimental data will be shown in Fig. 7).

For eliminating the heat loss through gas convection one simply needs a high vacuum. For eliminating radiation heat loss, however, simply using a radiation shielding at the substrate temperature $T_0$ will be helpless, because it is the radiation power inhomogeneity along the specimen that matters. Nevertheless, we feel that a simple heat shielding at $T_0$ will at least help minimizing the static radial heat current from the specimen to the environment, especially for measurements performed above room temperature. Otherwise such heat current could cause the temperature of the specimen inaccurate and the whole heat conduction processes complicated.

## VI. EXPERIMENTAL TESTS AND TIPS

We have tested this $3\omega$ method on two kinds of specimens: platinum wires and bundles of multiwall carbon nanotubes, by just using the approximation solution (19). The electrical resistance of the former specimen has a positive temperature coefficient and the latter a negative one. Within appropriate ranges of frequency and current, we do find that $V_{3\omega} \propto I^3$ and $V_{3\omega} \propto 1/\sqrt{1+(2\omega\gamma)^2}$. For the platinum specimen, the apparent specific heat and thermal conductivity as well as the Wiedemenn-Franz ratio agree with the standard data over the entire temperature range measured (10–320 K).

Figure 4 shows the block diagram for the measurement. A digital lock-in amplifier such as SR830 or SR850 made by Stanford Research Inc. was selected. All the filters on the lock-in were turned off, and the $dc$ coupled input mode was selected, to ensure the observation of a true frequency dependence of $V_{3\omega}$. Before measuring the $3\omega$ signal the phase of the lock-in amplifier was adjusted to zero according to the $1\omega$ voltage component. The phase angle of $V_{3\omega}$ is then $-\phi$ if $R' < 0$ or $180^o - \phi$ if $R' > 0$ according to (17). We used a simple electronic circuit (the lower panel of Fig. 4) to convert the $1\omega$ sine wave voltage from the sine-out of the lock-in to an $ac$ current, and then we fed the current into the specimen. The $3\omega$ component in the current was below $10^{-4}$ compared to its $1\omega$ component, checked by an HP89410A spectrum analyzer. Because the $3\omega$ voltage signal is deeply buried in the $1\omega$ voltage signal, certain amount of dynamic reservation is required for the lock-in if, in order to keep the simplicity of this method, not using a bridge circuit to cancel out the $1\omega$ signal. We kept the dynamic reservation unchanged relative to the total magnification of the lock-in during the entire measurement.

There are two ways to perform the measurement. In the first, the substrate of the specimen is maintained at fixed temperatures, then the frequency dependence of $V_{3\omega}$ is measured. In this way we can check the $I^3$ and the $1/\sqrt{1+(2\omega\gamma)^2}$ dependencies of $V_{3\omega}$ as well as the relation $\tan\phi = 2\omega\gamma$.

Because $V_{3\omega} \propto I^3$, one will get a much larger signal by using a larger $I$. However, there are three reasons for not using a very large $I$. First, it is required by condition (10). Second, radiation heat loss will be significant when the temperature modulation is large, as condition (31) tells. Third, excessive heat accumulation on the specimen would even create considerably large temperature gradient at the silver paste contacts, which might violate the boundary condition in (2). In all the cases the expected relations such as $V_{3\omega} \propto I^3$ will not be held. On the other side, the relation will also be violated if $I$ is too small so that $V_{3\omega}$ becomes comparable to, or even smaller than the spurious $3\omega$ signals that comes from the current or other sources. In our measurement, the total heating power was maintained such that the temperature modulation along the specimen was around 1 K. Nevertheless, if the $3\omega$ voltage is too small to measure then one has to increase the current for creating a larger temperature fluctuation. In this case the actual (averaged) temperature of the specimen has to be corrected afterwards by comparing the resistance of the specimen measured with the larger current and that measured with a much smaller one.

From (21), a longer and thinner specimen also gives a larger signal. However, a larger $L$ corresponds to a larger thermal time constant $\gamma$ ($\gamma \propto L^2$), and hence a lower frequency window for measurement. In practice, it will be inconvenient to perform the measurement below 1 Hz. A larger length and a smaller cross section or diameter could also violate the conditions (10) and (31), and thus violate the expected $I^3$ and the $1/\sqrt{1+(2\omega\gamma)^2}$ dependencies of $V_{3\omega}$.

In the second way of measurement, the temperature of the substrate is slowly ramped up or down at a fixed rate, meanwhile the working frequency of the lock-in amplifier is switched between a few set values. The maximum working frequency is adjusted by keeping $2\omega\gamma < 4$ (*i.e.*, $\phi < 76^o$ according to (18)). And the electric current is adjusted roughly to maintain a fixed $dc$ temperature accumulation (*i.e.*, $\sim 1$ K). The whole process including the temperature ramping, parameters adjusting, and frequency switching are all controlled by a personal computer.

For platinum specimen, we chose a wire of diameter $D$=20 $\mu$m and length $L$= 8 mm. We found that the thermal time constant $\gamma$ of the specimen varied from 0.005 s$^{-1}$ at 10 K to $\sim 0.2$ s$^{-1}$ at room temperature, so that the working frequencies were chosen to be between 1 to 80 Hz. Shown in Fig. 5 (a) is the current dependence of $V_{3\omega}$ at 25K, demonstrating an $I^3$ dependence in a mediate current range. Figure 5 (b) and (c) show the frequency dependencies of the amplitude and the phase angle of $V_{3\omega}$ at 25 K, compared with the predicted functional forms (the solid lines). By fitting the data in Fig. 5 (b) to (19), we obtained the thermal conductivity $\kappa$ (Fig. 5 (d), open circles) and the thermal time constant $\gamma$. The thermal diffusivity and the specific heat of the specimen can be obtained by using the relations $\gamma = L^2/\pi^2\alpha$ and $\alpha = \kappa/\rho C_p$. The results are shown in Fig. 5 (e) and (f) as open circles. $C_p$ thus obtained agrees well with the standard data for platinum[12] (the solid squares in Fig. 5 (f)).

The thermal conductivity of our platinum wire shows a less pronounced peak at low temperatures compared to that of high purity platinum. Since $\kappa$ depends largely on the purity, structural perfection, and even the size of the specimen, we think that the $\kappa$ data we obtained reflect the true thermal conductivity of our platinum wire. In fact, the Wiedemenn-Franz ratio of the specimen deduced from the thermal conductivity and the electrical resistivity, or more directly, deduced from the thermal conductance and the electrical resistance, fits to the case of pure but not totally defect-free metals[13], as shown in Fig. 6. The Wiedemenn-Franz ratio is found to be $\sim 2.53 \times 10^{-8} \text{W}\Omega/\text{K}^2$ at 290 K. It is slightly larger than the free-electron Lorenz number $2.45 \times 10^{-8} \text{W}\Omega/\text{K}^2$, and is rather close to $2.6 \times 10^{-8} \text{W}\Omega/\text{K}^2$, the reported value in literature for platinum[14].

Let us now examine the effect of radial heat loss through gas convection. The data in Fig. 5 were taken in a high vacuum where such heat loss was virtually absent, as that changing the vacuum pressure by a factor of 2 yielded a same $\kappa$. Shown in Figs. 7 (a), (b) and (c) are two sets of data taken on another platinum specimen at two different vacuum pressures. The circles represent the data taken in a vacuum where radial heat loss emerged but was not severe (indicated by the slightly positive slope of $\kappa$ at at high temperatures). During one of the warming-up measurements, however, we introduced radial heat loss by destroying the system's vacuum. After that, spurious larger thermal conductivity and diffusivity of the specimen were obtained, shown as the squares in Figs. 7 (a) and (b)). The radial heat current reached several times larger than the axial one at room temperature, as indicated in Fig. 7 (a). Nevertheless, the specific heat deduced from $\kappa$ and $\alpha$ was quite insensitive to the radial heat loss (Figs. 7 (c)). The reason has been explained in section V.

After all, let us check if conditions (10) and (31) were satisfied. If taking $n = 1$, we had $\frac{I_0^2 R'L}{n^2\pi^2\kappa S} \sim 10^{-5}$. Therefore condition (10) was well satisfied. For condition (31), assuming an emissivity $\epsilon = 1$ for our platinum wire leads to $g \approx 0.44 \text{ s}^{-1}$ at 300 K. On the other hand, $\gamma$ (actually, $\gamma_{ap}$) deduced from the measurement was $\sim 0.2$ s. Therefore, $g\gamma \approx 0.088$. In the real case the product $g\gamma$ should be much smaller than 0.088, because the emissivity of a shiny metal is usually much less than unity. Therefore, condition(31) should also be well satisfied.

We have also applied the $3\omega$ method to measure the $\kappa$ and $C_p$ of multiwall carbon nanotube (MWNT) bundles who have a negative $R'$ (Ref. 15). MWNT is a highly anisotropic material both in geometry and in thermal conductivity, owing to its strong in-plane $sp^2$ bonding and the weak interwall van der Wasls bonding. Its macroscopic length against nanometer-sized diameter ensures overall a much shorter thermal time constant in the radial direction than in the axial direction. We believe this conclusion is also true for a bundle of MWNTs. Therefore the heat conduction can be regarded as a 1D problem. For MWNTs there is no $C_p$ and $\kappa$ data of other sources available for comparison. Nevertheless, the obtained frequency and current dependencies of $V_{3\omega}$ were all in good agreement with (19) (Fig. 8), which guarantees the reliability of $\kappa$ and $C_p$ thus obtained. For a carbon nanotube bundle of $L = 1$ mm and $D = 10\mu$m, $\frac{I_0^2 R'L}{\pi^2\kappa S}$ was less than $10^{-3}$ at temperatures above 60 K, and was about 0.08 at 10 K. In addition, the product $g\gamma$ was below $4 \times 10^{-3}$ in the whole temperature range investigated (estimated using the emissivity of a black body). Therefore, both conditions (10) and (31) were satisfied if considering the bundle as a unitary object. The nanotubes inside the bundle were actually "self-shielded" by the outmost ones if examining them individually, which might effectively eliminate the radial heat loss.

For a carbon nanotube bundle of 1 $\mu$m in diameter and 1 mm in length, its mass is only around $10^{-9}$ g, far less than the minimum amount of mass (typically in mg) required in many other kinds of $C_p$ measurement.

## VII. CONCLUSION

We have explored a $3\omega$ method for measuring the thermal conductivity and specific heat of a rod or filament-like specimen. By fitting the frequency-dependent $3\omega$ voltage data to (19) within the frequency range $0 < 2\omega\gamma < 4$, we can obtain $\kappa$ and $C_p$ of the specimen to an accuracy of 2-4%. For achieving higher accuracy one can fit the data to (24). The presence of radial heat loss will result in a larger apparent thermal conductivity. But $C_p$ obtained by this method is very much insensitive to such heat loss and thus keeps to be reliable. A successful measurement relies on properly choosing the specimen's dimensions, so that one can have a large enough $3\omega$ voltage, yet maintaining a convenient working frequency range and keeping the condition (10) (and (31) if necessary) satisfied.

## ACKNOWLEDGMENTS

We thank G. H. Li and X. N. Jing for helpful discussion. This work was supported by NSFC, and by the President's Foundation of CAS.

FIG. 1. Illustration of the four-probe configuration for measuring the specific heat and thermal conductivity of a rod- or filament-like specimen. The specimen is heat sunk to the sapphire substrate through the four electric contacts, but the part in-between the two voltage contacts needs to be suspended, to allow the temperature variation. A high vacuum is needed and a thermal shielding is preferred to eliminate the radial heat current from the specimen to the environment.

FIG. 2. Temperature fluctuation along the specimen driven by an *ac* current $I_0 \sin\omega t$. The fluctuation amplitude is marked as shadowed area. It reaches the maximum at the limit $\omega\gamma \to 0$, and shrinks to a line as $\omega\gamma \to \infty$. The line in the middle of the fluctuation range denotes the *dc* temperature accumulation along the specimen, which reaches the maximum value of $\Delta_0$ (defined in the text) at the center of the specimen.

FIG. 3. The errors of $V_{3\omega}$ caused by truncating the $n>1$ terms in (16). Curve A represents the exact solution of the $3\omega$ voltage amplitude. Curve B is the $3\omega$ voltage of the $n=1$ term alone. The difference between them is nearly a constant at low frequencies, plotted as curve A-B. The relative error of $V_{3\omega}$ increases with $\omega$, illustrated as curve (A-B)/A.

FIG. 4. Block diagram of the measurement. A digital lock-in amplifier such as SR830 or SR850 was chosen to measure the $3\omega$ voltage. The $1\omega$ voltage from the sine-out of the lock-in was boosted into an *ac* current by a simple electronic circuit (lower panel), and was then fed into the specimen. The feed-back resistor R* should have a nearly temperature-independent coefficient to prevent from generating a $3\omega$ component in the current.

FIG. 5. Experimental test of the $3\omega$ method on a platinum wire of 20 $\mu$m in diameter. (a) The current dependence of $V_{3\omega}$. The open circles are the measured data at 25 K and 2 Hz, and the solid line is the predicted relation $V_{3\omega} \propto I^3$. (b) The frequency dependence of $V_{3\omega}$ at 25 K (open circles). The solid line is the predicted relation $V_{3\omega} \propto 1/\sqrt{1+(2\omega\gamma)^2}$. (c) The frequency dependence of the phase angle of $V_{3\omega}$ at 25 K (open circles). The obtained thermal conductivity $\kappa$, thermal diffusivity $\alpha$, and specific heat $C_p$ of the platinum specimen are plotted as open circles in figures (d), (e), and (f), respectively. Also shown in (d), (e), and (f) as solid squares are the standard data of platinum from literature[12]. The difference in $\kappa$ and $\alpha$ between our data and the standard ones should reflect the difference in purity and/or structural perfection between the platinum specimens of different sources.

FIG. 6. The Wiedemenn-Franz ratio $L^e$ of the platinum specimen compared with that of usual metals with different purity. The result indicates that the platinum wire used in this experiment is pure but not totally defect-free. The room temperature Wiedemenn-Franz ratio of the platinum wire is about $2.52 \times 10^{-8}$W$\Omega$/K$^2$, which is in good agreement with the reported value of $2.6 \times 10^{-8}$W$\Omega$/K$^2$ in literature[14]. The Lorenz number of free electron gas is $L_0 = 2.45 \times 10^{-8}$W$\Omega$/K$^2$, plotted as the dashed line. Note that for platinum the Debye temperature $\theta$ is 240 K.

FIG. 7. Effect of radial heat loss through air convection. The circles represent the data taken in a vacuum where radial heat loss was not significant. In one warming-up run of the measurement, radial heat loss was triggered on above $T^*$ by destroying the system's vacuum. The heat loss resulted in a spurious larger thermal conductivity and diffusivity for the specimen (the squares in (a) and (b)). But, as predicted by (33), the specific heat deduced from them was relatively insensitive to such heat loss (the squares in (c)).

FIG. 8. Experimental test of the $3\omega$ method on multiwall carbon nanotube bundles at 50 K. (a) The current dependence of the $3\omega$ voltage measured at 10 Hz compared with the predicted form $V_{3\omega} \propto I^3$ (the solid line). (b) The frequency dependence of $V_{3\omega}$ compared with the predicted relation $V_{3\omega} \propto 1/\sqrt{1+(2\omega\gamma)^2}$. (c) The frequency dependence of the phase angle of $V_{3\omega}$ compared with the predicted relation $\tan\phi \propto \omega$. The temperature dependencies of the thermal conductivity, thermal diffusivity and specific heat of the material have already been published elsewhere[15].

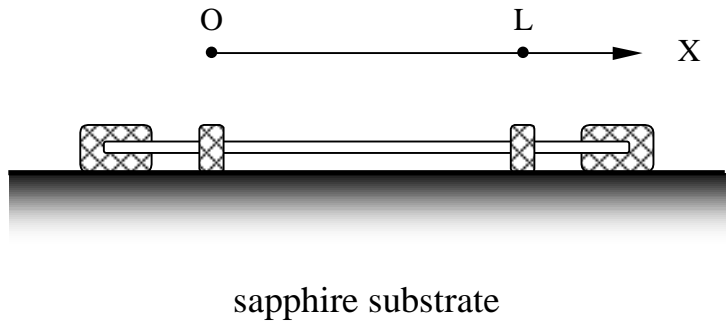

sapphire substrate

L. Lu, et al., Fig. 1

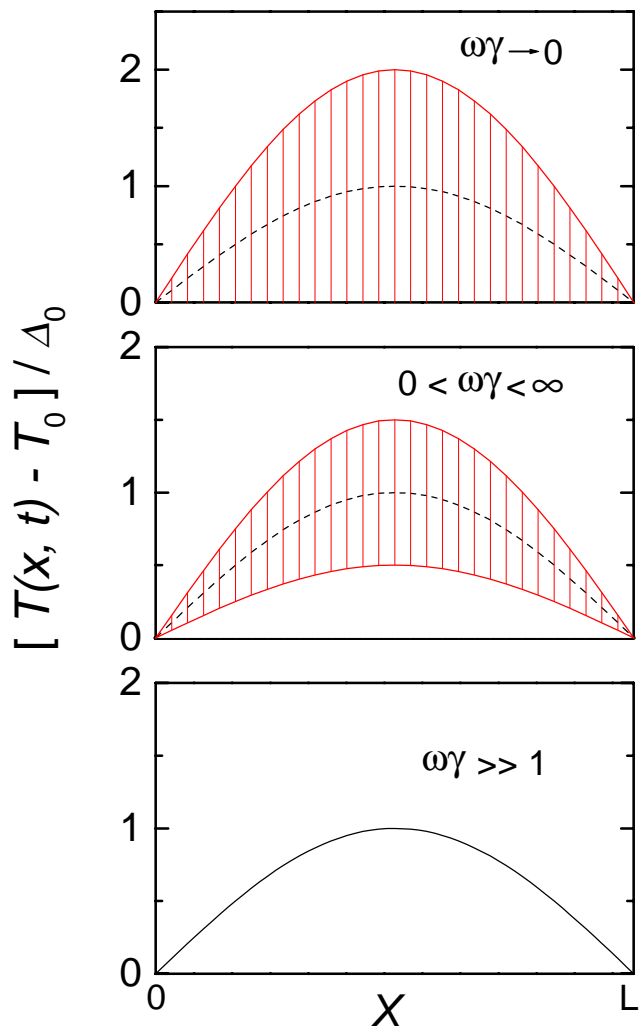

L. Lu, et al., Fig. 2

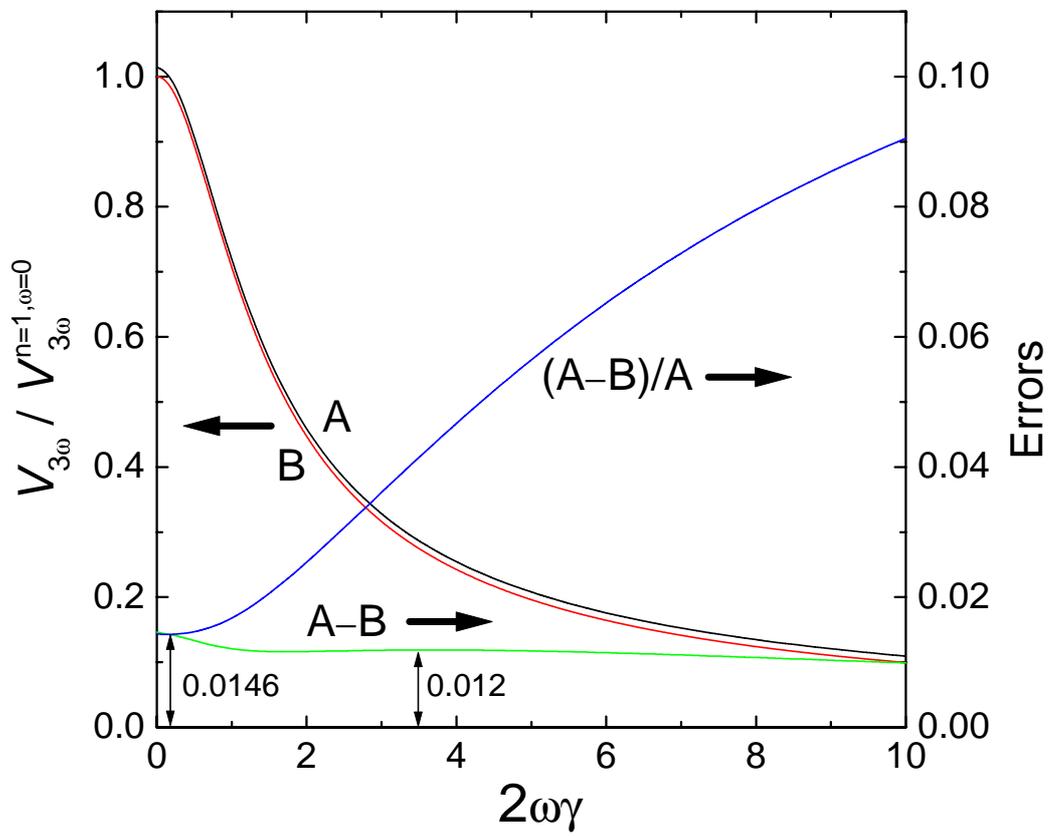

L. Lu, et al., Fig. 3

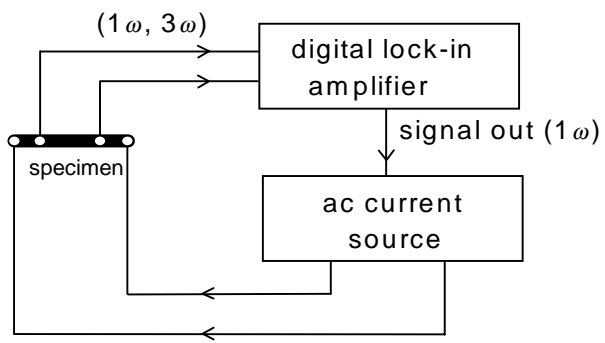
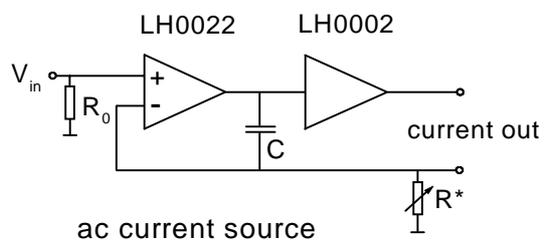

L. Lu, et al. Fig. 4

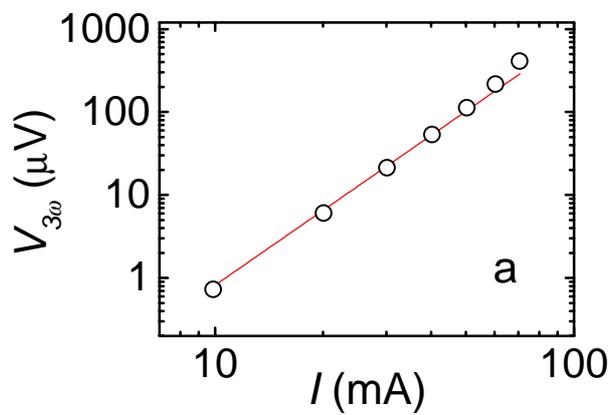
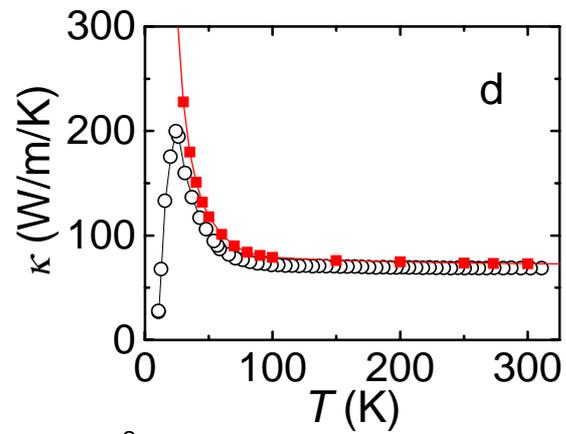
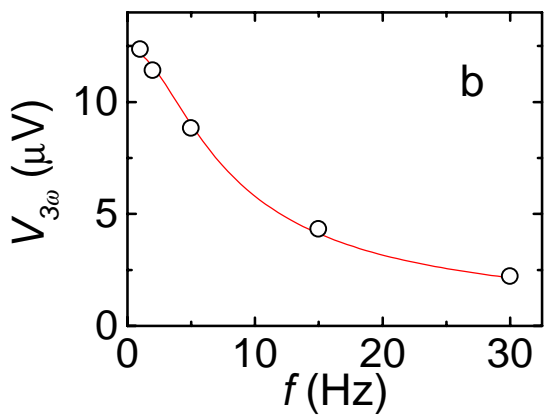
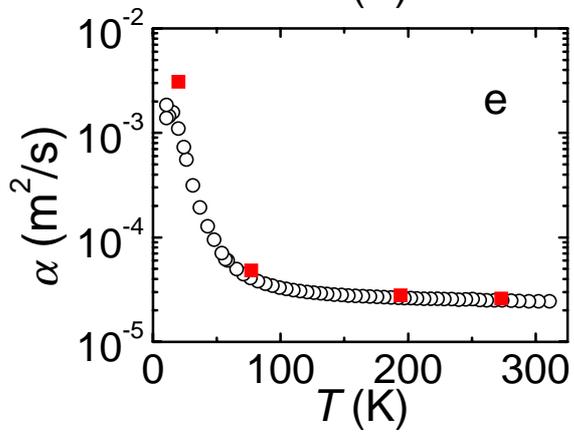
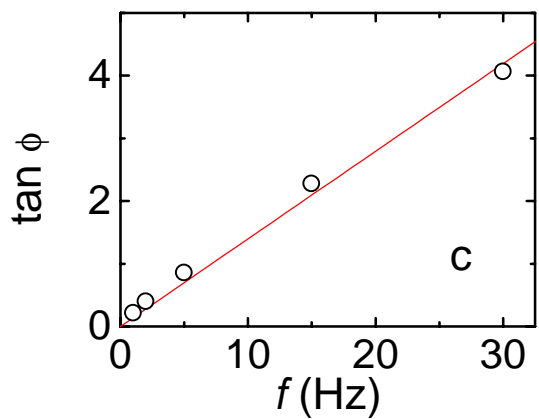
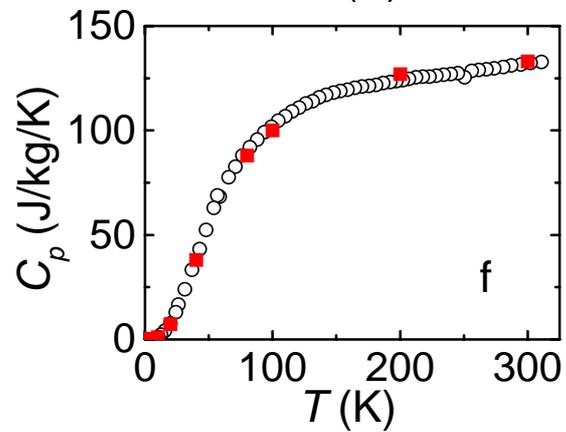

L. Lu et al.  Fig. 5

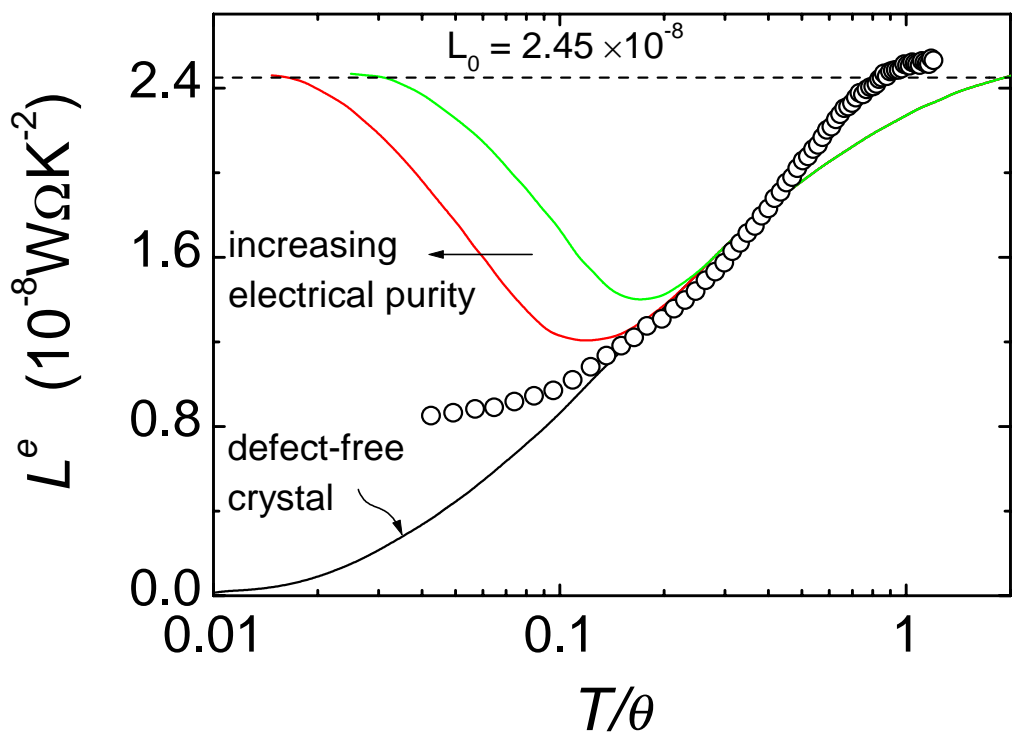

L. Lu et al. Fig. 6

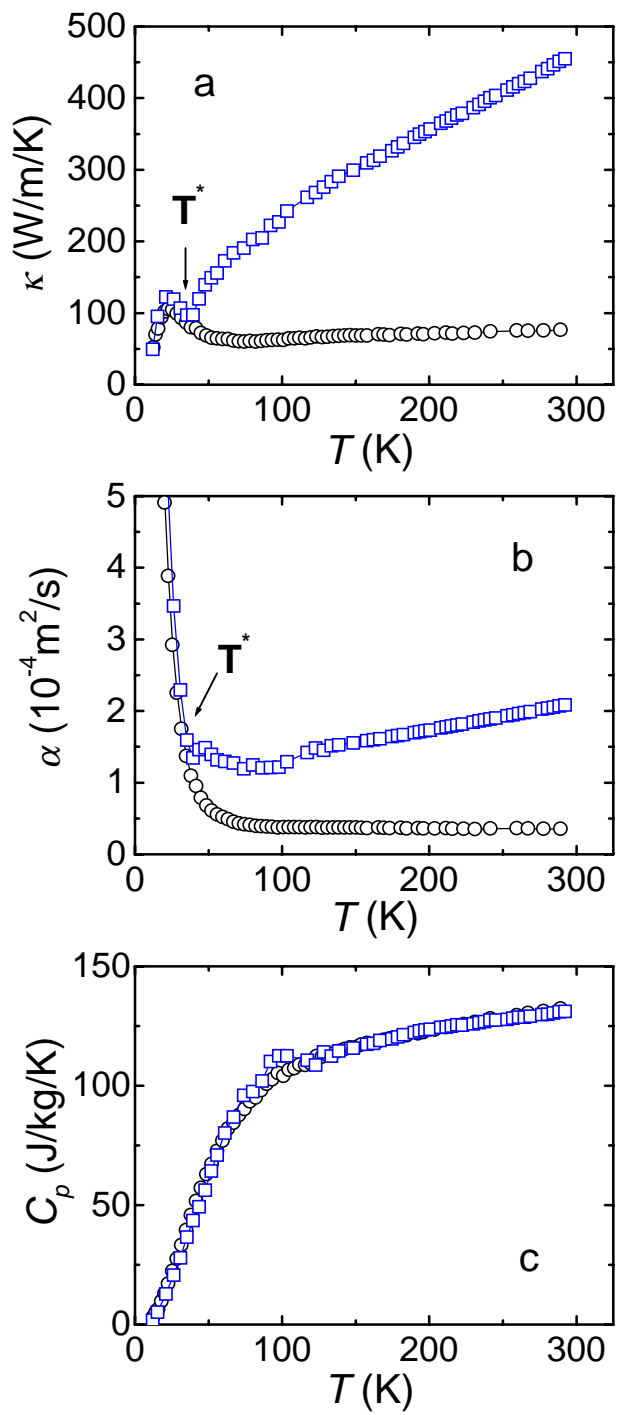

L. Lu et al. Fig. 7

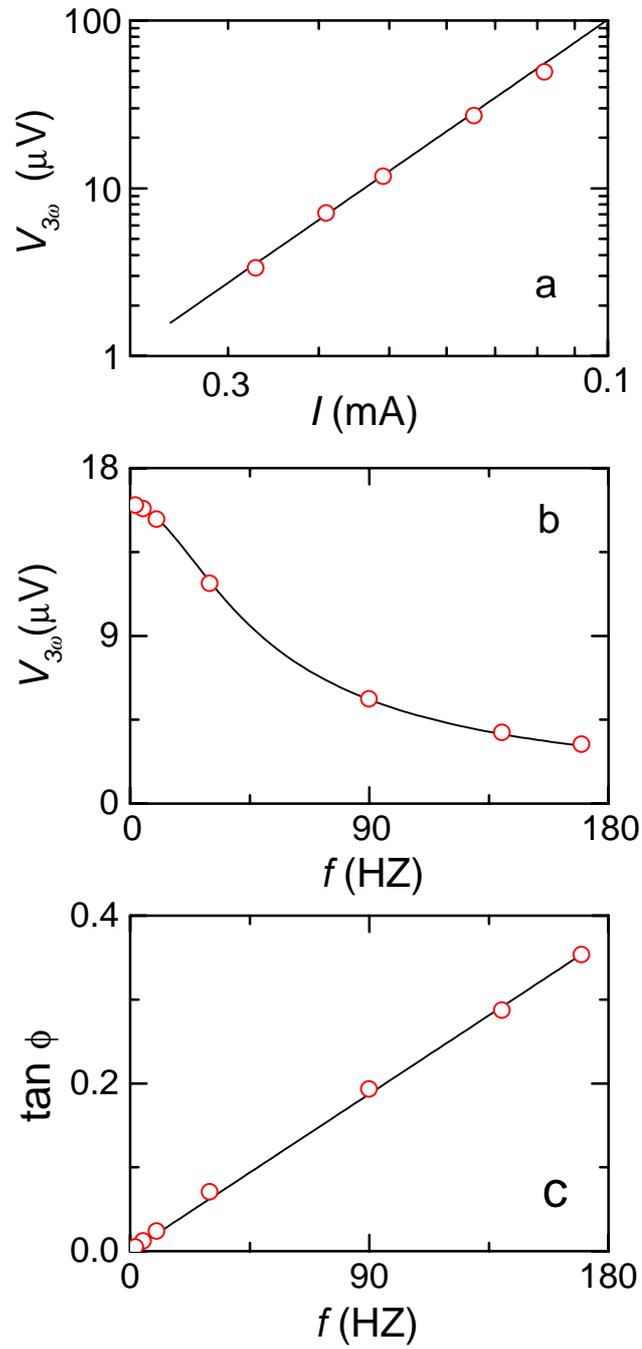

L. Lu, et al., Fig. 8